\documentclass[twocolumn, amsmath,nobibnotes,nofootinbib,showpacs,aps,prb]{revtex4}  
\usepackage[dvips]{graphicx}
\begin{document}
\title{Possible magnetoelectric coupling in the half doped charge ordered manganite, Pr$_{0.5}$Ca$_{0.5}$MnO$_{3-\delta}$}
\author{A. Karmakar$^1$}
\author{S. Majumdar$^1$}
\author{A. K. Singh$^2$}
\author{S. Patnaik$^2$}
\author{S. Giri$^1$}
\email{sspsg2@iacs.res.in} 
\affiliation{$^1$Department of Solid State Physics, Indian Association for the Cultivation of Science, Jadavpur, Kolkata 700032, India \\$^2$School of Physical Sciences, Jawaharlal Nehru University, New Delhi 110067, India}

\begin{abstract}
Magnetization, magnetoresistance, and magnetodielectric measurements have been carried out on the half doped charge ordered manganite, Pr$_{0.5}$Ca$_{0.5}$MnO$_3$. The low temperature state is found to be strongly dependent on the oxygen stoichiometry whereas the high temperature state remains almost unaltered. A disorder driven phase separation in the low temperature state is noticed in the magnetic, magnetoresistance, and dielectric measurements which is attributed to the oxygen deficiency in the compound. A considerable magnetodielectric (MD) effect is noticed close to room temperature at 280 K which is fascinating for technological applications. The strongest MD effect observed in between 180 K and 200 K is found to be uncorrelated with magnetoresistance but it is suggested to be due to a number of intricate processes occuring in this temperature range which includes paramagnetic to antiferromagnetic transition, incommensurate to commensurate charge ordering and orbital ordering. The strongest MD effect seems to emerge from the high sensitivity of the incommensurate state to the external perturbation such as external magnetic field. The results propose the possible magnetoelectric coupling in the charge ordered compound.
\end{abstract}


\maketitle

\section{Introduction}
%capacitors, transducers, actuators, etc
Materials in which coupling between magnetic and electric order exists, are typically defined as multiferroics or magnetoelectrics. \cite{fie,che} Such single-phase materials are relatively very  rare, which arises the fundamental question about the possible mechanisms for multiferroic behavior. \cite{khom1,eere} Coupling between magnetic and electric polarization in principle has tremendous application in writing electrically and reading magnetically (and vice versa) which is attractive for exploiting it toward a new generation of spintronic devices such as ferroelectric random access memory (FRAM), magnetic data storage, etc. 

Magnetoelectric coupling can be measured indirectly by simply recording changes in either magnetization near a ferroelectric transition or the dielectric constant near a magnetic transition temperature. The resulting effects are typically defined as magnetocapacitance or magnetodielectric effect. Catalan \cite{catalan} has pointed out that  magnetodielectric or magnetocapacitance effect may be achieved through a combination of magnetoresistance and the Maxwell-Wagner effect which is not the intrinsic magnetoelectric coupling, despite the fact that this resistive magnetocapacitance has tremendous technological applications like a magnetic field tunable resonant frequency in a capacitive resonator. Recently, strong magnetoelectric effect at the (ferro/antiferro)magnetic transition has been reported in the magnetic dielectrics such as $R$MnO$_3$ ($R$ = Gd, Tb, Dy) \cite{kim,cheo1,kim1,senff} and  $R$Mn$_2$O$_5$ ($R$ = Tb, Dy, Ho), \cite{cruz,blake,hur} BiMnO$_3$, \cite{santos,kimura2} BiFeO$_3$ \cite{wang,singh}  where different origins of intrinsic magnetoelectric coupling have been discussed in the literatures. 

Recent reviews \cite{brink,cnr} propose that charge ordering (CO) in the mixed valent manganites with perovskite structure might be a new paradigm for multiferroicity where a beautiful compromise between the two extremes, the site-centered charge ordering (SCO) and the bond-centered charge ordering (BCO) in manganites may give rise to the ferroelectricity. \cite{brink,efre} The presence of BCO and SCO in Pr$_{1-x}$Ca$_x$MnO$_3$ (0.3 $\leq x \leq$ 0.5) has been pointed out in recent reports depending on the hole doping where these compounds have been proposed as potential candidates for ferroelectricity. \cite{brink,efre,daou,van} Lopes {\it et al}. clearly demonstrated that temperature dependence of electric field gradiant (EFG) associated with polar atomic vibrations leads to a spontaneous local electric polarization below CO transition in Pr$_{1-x}$Ca$_x$MnO$_3$ (0.25 $\leq x \leq$ 0.85). \cite{lopes} A direct observation of ferroelectricity in Pr$_{1-x}$Ca$_x$MnO$_3$ could settle the case. Unfortunately, the development of a macroscopic polarization is hindered by a relatively high conductivity of this system, also preventing direct measurement of the polarization. Till date, a peak at the CO temperature was reported in the dielectric constant for Pr$_{1-x}$Ca$_{x}$MnO$_3$ with $x$ = 0.33 \cite{jard,merc} and 0.40 \cite{bisk1} suggesting the indication of a magnetoelectric coupling. Lately, large magnetodielectric response has been observed in $x$ = 0.30, \cite{frei,kund} and 0.40,  \cite{bisk2} although the origin of magnetoelectric coupling remains controversial. 

In this article, we report a fairly large magnetodielectric (MD) response in the half doped charge ordered compound, Pr$_{0.5}$Ca$_{0.5}$MnO$_3$, even near room temperature, which has tremendous technological applications. Largest effect in the MD response is observed in between 180 K and 200 K around which several intricate phenomena associated with the incommensurate to commensurate (IC to C) charge and orbital ordering transition take place along with a paramagnetic to antiferromagnetic transition ($T_{\rm N}$) at 180 K. \cite{tomi,mori,asa,kaji,zim,yaku} The results are significantly analogous to that observed in multiferroic,  $R$Mn$_2$O$_5$ ($R$ = Tb, Dy, Ho) where colossal MD effect was correlated to the C to IC magnetic transition. \cite{cruz,blake,hur} We further note that the strongest MD effect around $T_{\rm N}$ is not a consequence of magnetoresistance. Hence, a possible magnetoelectric coupling in Pr$_{0.5}$Ca$_{0.5}$MnO$_3$ rises a fundamental question whether the compound can be recognized as an elite member in multiferroics.  

\section{Experimental}
Polycrystalline compound, Pr$_{0.5}$Ca$_{0.5}$MnO$_3$ was prepared by the standard sol-gel technique. \cite{KDe} Pr$_2$O$_3$,  CaCO$_3$, and Mn(CH$_3$COO)$_2$,4H$_2$O were used as starting materials. The final heat treatments were performed at 1000$^\circ$C for 12 h in air. In order to observe the effect of oxygen stoichiometry on the physical properties a part of the sample was annealed at 1000$^\circ$C for 15 h in the atmospheric pressure of oxygen gas. For simplicity, we address the sample annealed in open air as oxygen deficient and that annealed in closed oxygen atmosphere as oxygen annealed sample. The average grain sizes were $\sim$ 0.3 $\mu$m obtained from Scanning Electron Microscopy using a microscope  (JSM-6700F, JEOL). The single phase of the crystal structure was confirmed by a BRUKER axs x-ray powder diffractometer (8D - ADVANCE). Magnetization was measured using a  commercial superconducting quantum interference device (SQUID) magnetometer (MPMS, XL) of Quantum Design. Capacitance ($C$) and ac conductance ($G$) measurements were carried out using an LCR meter (E4980A, Agilent). Electrical contacts of the circular disc shaped samples were fabricated using air drying silver paint. Magnetocapacitance and magnetoresistance measurements were carried out using a commercial superconducting magnet system (Cryogenic Ltd., UK). Resistivity was measured using a standard four probe technique. Heat capacity as a function of temperature ($T$) was measured using a home built set up. We use zero-field cooled heating (ZFCH), field-cooled heating (FCH), and field-cooled cooling (FCC) modes for the magnetization, magnetoresistance, and dielectric measurements. In case of ZFCH mode the sample was cooled down to the lowest temperature in zero field and measurement was performed with field in the heating cycle while for FCH mode the sample was cooled in field and measured in the warming cycle with the field applied. The measurement in FCC mode was carried out in the cooling cycle with an applied magnetic field. 

\begin{figure}[t]
\vskip 0.0 cm
\centering
\includegraphics[width = 8 cm]{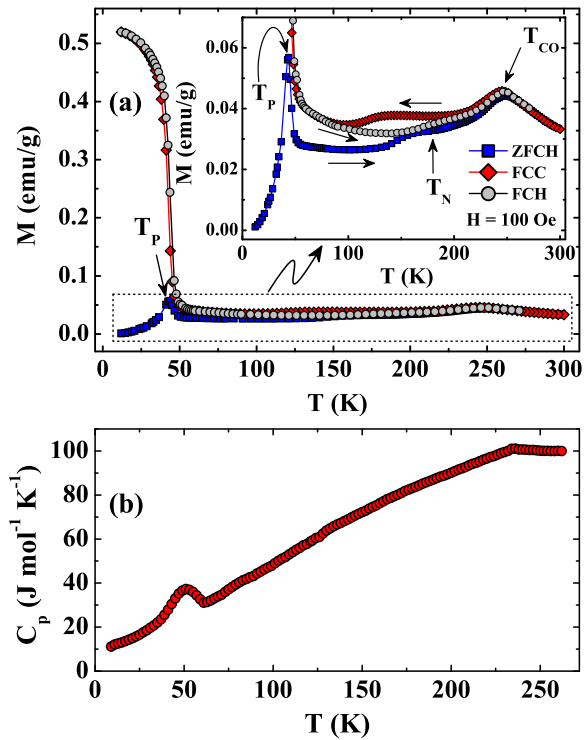}
\caption {(a) Magnetization ($M$) as a function of temperature ($T$) in the ZFCH, FCH and FCC modes at field, $H$ = 100 Oe. Inset highlights the peaks at $T_{\rm P}$ and $T_{\rm CO}$ in an expanded scale. (b) Heat capacity ($C_{\rm P}$) as function of $T$ for the oxygen deficient sample.} 
\label{Fig. 1}
\end{figure}
\section{Results and discussions}
Temperature dependence of dc magnetization measured at 100 Oe with different modes is shown in Fig. 1(a) for the oxygen deficient sample. At high temperature a peak at 245 K corresponding to the CO temperature ($T_{\rm CO}$) and a step like feature below $\sim$ 180 K ($T_{\rm N}$) are noticed in the ZFCH magnetization in accordance with the reported results. \cite{tomi} The field-cooled (FC) effect of magnetization is observed below $T_{\rm CO}$. Thermal hysteresis between FCH and FCC magnetization is highlighted below $T_{\rm CO}$ in the inset of Fig. 1(a) which is typical for charge ordered compounds, exhibiting the signature of first order transition at $T_{\rm CO}$. We further note that $T_{\rm N}$ indicated by the step in the FCC curve is shifted toward lower temperature around $\sim$ 140 K whereas in the FCH curve, a weak anomaly can be noted toward higher temperature around $\sim$ 200 K. The results clearly demonstrate that microstructures of C and IC charge ordered states coexist in the thermally irreversible temperature region, below $T_{\rm CO}$, in accordance with the reported results. \cite{mori,asa,kaji,zim,yaku} With further decreasing temperature a sharp peak ($T_{\rm P}$) at 43 K is  noticed with an extraordinarily large FC effect below $T_{\rm P}$. The sharp peak at $T_{\rm P}$ is highlighted in the inset of Fig. 1(a). The results are in accordance with the previous reports. \cite{doerr,cao} Heat capacity ($C_{\rm p}$) as a function of temperature is shown in Fig. 1(b) where a peak and an anomaly are observed around $T_{\rm P}$ and $T_{\rm CO}$, respectively. Recently, a strong magnetoelastic coupling has been proposed in Pr$_{0.5}$Ca$_{0.5}$MnO$_3$ where an anomalous lattice shrink ($\Delta l/l \sim 1.5 \times 10^{-3}$) below $T_{\rm P}$ was reported by Doerr {\it et al}. \cite{doerr} The peak observed at $T_{\rm P}$ in $C_{\rm p}$ is consistent with the anomalous lattice shrink and confirms that $T_{\rm P}$ is not a reentrant spin-glass transition temperature proposed by Cao {\it et al}. \cite{cao} However, for the oxygen annealed sample, the peak height at $T_{\rm P}$ is enormously decreased but the peak at the charge order transition $T_{\rm CO}$ remains more or less the same, the peak positions being unchanged in both the cases. Possible origin of $T_{\rm P}$ has been discussed below elsewhere in the text. 

Magnetization curves in the increasing and decreasing cycles of field were measured above and below $T_{\rm P}$ at 70 and 20 K, for the oxygen deficient sample, which are displayed in Fig. 2(a) and (b), respectively. Immeasurable remanence and coercivity for the magnetization curve at 70 K indicate that the ferromagnetic component is not developed even for an applied magnetic field up to 50 kOe in accordance with a previous report. \cite{toku} On the other hand at 20 K, a small remanence is noticed even for the measurement up to 10 kOe and the remanence remains almost unchanged for the measurement up to 30 kOe above which a sharp increase in magnetization is observed. This sharp increase appears due to the development of ferromagnetic component by destroying the charge ordered state which is demonstrated by the minor loop effects above 30 kOe. The remanence is considerably enhanced for the measurements with magnetic field above 30 kOe where remanence increases with the increase in the applied magnetic field. Interestingly, a similar magnetization curve for the oxygen annealed sample [Fig. 2(c)] measured at a temperature of 10 K, which is well below $T_{\rm P}$, shows a negligible remanence and coercivity. This indicates that for the oxygen stoichiometric sample the fraction of the ferromagnetic component, which develops below $T_{\rm P}$ at the cost of the charge ordered component, is negligible. Thus, it's obvious that the disorder originating form a non-stoichiometry of oxygen mainly drives the intricate processes at $T_{\rm P}$.

\begin{figure}[t]
\vskip 0.0 cm
\centering
\includegraphics[width = 8 cm]{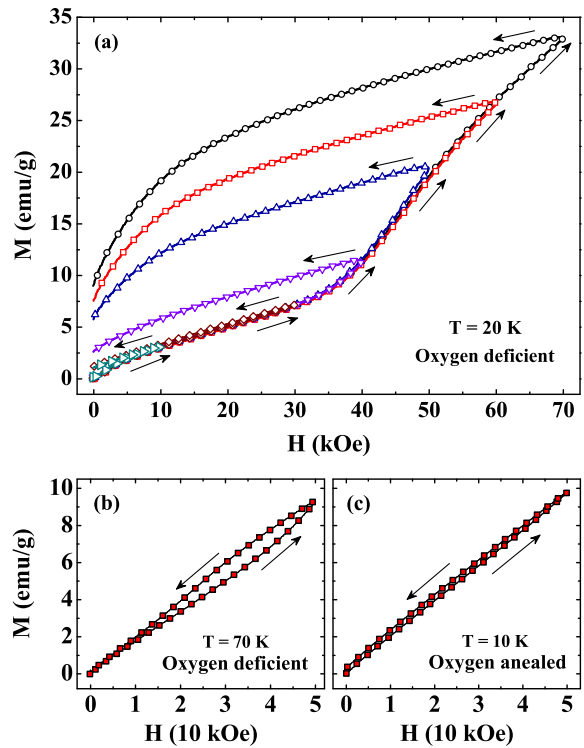}
\caption {(a) Minor loop effect of the magnetization ($M$) curves as a function of magnetic field ($H$) at 20 K for 6 different maximum magnetic field values {\it viz}, 10, 30, 40, 50, 60 and 70 kOe for the oxygen deficient sample. The directions of field cycling are shown by arrows. (b) and (c) show the $M$($H$) curves at 70 K for the oxygen deficient sample and at 10 K for the oxygen annealed sample, respectively.} 
\label{Fig. 2}
\end{figure}

In accordance with the magnetic results below $T_{\rm P}$, resistivity measured at 50 kOe  exhibits a semiconducting to metallic like transition at 50 K, close to $T_{\rm P}$, which is shown in the inset of Fig. 4 for the oxygen deficient sample. The comparatively low resistive phase below 50 K is suggested due to the appearance of ferromagnetic metallic phase by the partial destruction of the charge ordered state. We note that resistivity ($\rho$) could not be measured at low temperature ($T <$ 80 K) in zero field and in 30 kOe field in the cooling cycle because of large magnitude of $\rho$. It could be measured only when a higher magnetic field of 50 kOe was applied simultaneously during the resistivity measurement. The results are in conformity with the magnetization curve at 20 K [Fig. 2(a)]. Ferromagnetic components are developed on the application of magnetic field above a critical value around $\sim$ 35 kOe. A considerable thermal hysteresis between the heating and cooling cycles is observed in the range 25 - 115 K for the measurement in 50 kOe field which is ascribed to the coexistence of ferromagnetic metallic and insulating charge ordered states. In the present observation the magnetization and magnetoresistance results are in accordance with the previously reported results \cite{toku} in a single crystal above $T_{\rm P}$ whereas it differs markedly below $T_{\rm P}$. The reported results indicate that at 4 K the charge ordered state destroys at much higher field around $\sim$ 150 kOe than that of the current investigation. In order to find out the origin of the difference we measured the magnetoresistance in the oxygen annealed sample. Like the oxygen deficient sample the oxygen annealed sample was cooled down to 5 K in a 50 kOe field. Nevertheless, $\rho$ could not be measured at low temperature because of the large value. We do not observe any feature like the oxygen deficient sample at low temperature measured in 50 kOe field which is rather consistent with the results observed in the single crystal. \cite{toku} In fact, the sharpness of the peak at $T_{\rm P}$ observed in the temperature variation of dielectric loss ($\tan\delta$) for the oxygen deficient sample decreases substantially due to the oxygen annealing [see Fig. 3(b)] which confirms that the strong evidence in magnetization, resistivity, and dielectric loss at $T_{\rm P}$ may be correlated to the oxygen deficiency in Pr$_{0.5}$Ca$_{0.5}$MnO$_3$. It may be suggested that oxygen non-stoichiometry leads to the disorder in the charge ordered state giving rise to the phase coexistence at low temperature. \cite{burgy,roz} However, a detailed understanding using microscopic experimental tools is required for the oxygen deficient sample to elucidate the underlying nature below $T_{\rm P}$. Herein, we would like to focus on the observed considerable magnetodielectric effect close to room temperature. 

\begin{figure}[t]
\vskip 0.0 cm
\centering
\includegraphics[width = 8 cm]{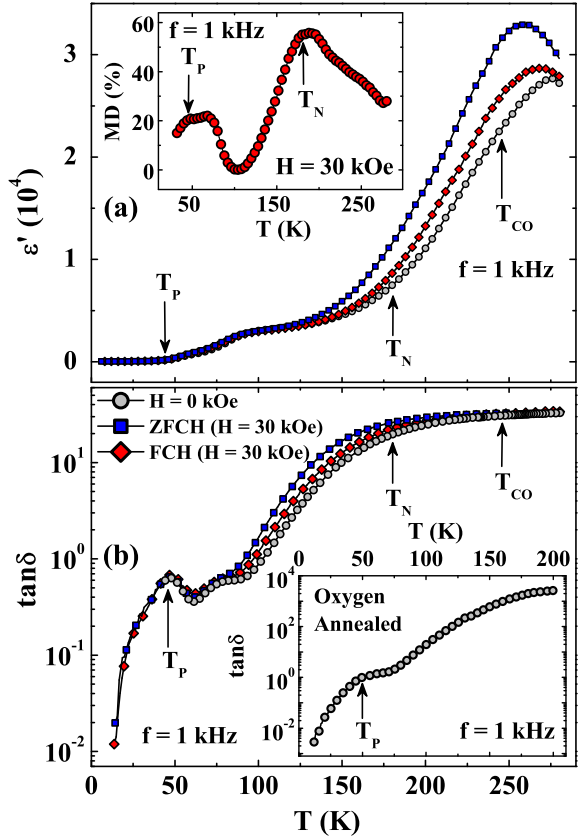}
\caption {(a) Temperature ($T$) variation of real part of dielectric permittivity ($\epsilon^\prime$) at $f$ = 1 kHz in the heating cycle in three different modes $viz$, without field, at $H$ = 30 kOe  with ZFCH and FCH modes for the oxygen deficient sample. Inset shows the variation of the magnetodielectric effect, MD(\%) defined by Eq. (1), as a function of temperature. (b) $T$-dependence of $\tan\delta$ at $f$ = 1 kHz in the heating cycle with zero field, ZFCH and FCH modes for the oxygen deficient sample. Inset shows the $T$ variation of $\tan\delta$ for the oxygen annealed sample at $f$ = 1 kHz and $H$ = 0 Oe.} 
\label{Fig. 3}
\end{figure}

\begin{figure}[t]
\vskip 0.0 cm
\centering
\includegraphics[width = 8 cm]{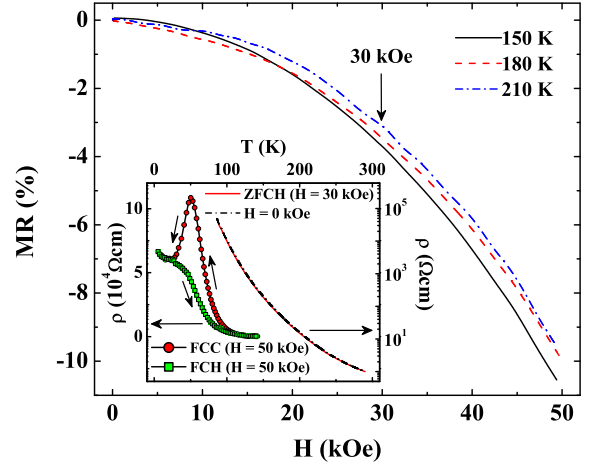}
\caption {Magnetic field ($H$) dependence of magnetoresistance (MR) at 150, 180, and 210 K, respectively. Inset shows the resistivity ($\rho$) as a function of temperature ($T$) in zero field, ZFCH and FCH modes with different fields. In the right axis $\rho$ at $H$ = 0 and in ZFCH mode with 30 kOe field are shown by the semilogarthmic plot. In the left axis $\rho$ in FCC and FCH modes with 50 kOe field are shown by the linear plot. All the results are given for the oxygen deficient sample.} 
\label{Fig. 4}
\end{figure}

We have measured the complex dielectric permittivity ($\epsilon$) as a function of temperature at frequency, $f$ = 1 kHz in different modes of applied magnetic field ($H$). Temperature variation of the real part ($\epsilon'$) of $\epsilon$ and the dielectric loss ($\tan\delta = \epsilon''/\epsilon'$) in zero-field, ZFCH, and FCH modes measured at 30 kOe are displayed in Figs. 3(a) and (b), respectively, for the oxygen deficient sample where $\epsilon''$ is the imaginary component of $\epsilon$. The strong magnetodielectric (MD) effect as well as the FC effect in $\epsilon'$ and $\tan\delta$ are observed through out the measured temperature region. Interestingly, the signature of $T_{\rm CO}$ is evident from $\tan\delta$($T$) where ZFCH, FCH, and the curve at $H$ = 0 merge at $T_{\rm CO}$ above which $\tan\delta$ shows nearly temperature and field independent results. We note that the signature of $T_{\rm CO}$ is not observed in $\epsilon'$ as well as in resistivity which are shown in Fig. 3 and inset of Fig. 4, respectively. A large positive MD effect displaying a considerable difference between the zero-field-$\epsilon'$ and the ZFCH-$\epsilon'$ at $H$ = 30 kOe is shown in the inset of Fig. 3(a). A quantitative estimate of MD effect can be obtained from the relation
\begin{equation}
\textup {MD} (\%) = \frac{\epsilon'(H)-\epsilon'(0)}{\epsilon'(0)} \times 100
\end{equation}
where $\epsilon'(H)$ is the dielectric permittivity measured in ZFCH mode with $H$ = 30 kOe and $\epsilon'(0)$ is that measured in $H$ = 0 kOe. The plot of MD (\%) with $T$ shows maxima around $T_{\rm N}$ and $T_{\rm P}$ with magnitudes of 56 \% and 22 \% at $T_{\rm N}$ and $T_{\rm P}$, respectively. In fact, a considerable MD effect with MD (\%) = 28 \% is also observed close to room temperature at 280 K. The strongest MD effect observed in between 180 - 200 K is significant where several transitions including the paramagnetic  to antiferromagnetic, IC to C charge and orbital ordering have been reported.  \cite{tomi,mori,asa} We note that the results are analogous to the strongest MD effect in multiferroic $R$Mn$_2$O$_5$ ($R$ = Tb, Dy, Ho) which is involved with an unusual commensurate to incommensurate magnetic transition having antiferromagnetic ground state. \cite{cruz,blake,hur} Recently, large magnetodielectric response has also been reported in Pr$_{1-x}$Ca$_{x}$MnO$_3$ with $x$ = 0.30, \cite{frei,kund} and 0.40, \cite{bisk2} despite the fact that the origin of magnetoelectric coupling remains controversial. One of the common interpretation of MD effect is due to considerable magnetoresistance. \cite{catalan}

Magnetoresistance as a function of temperature was measured at 30 kOe in which magnetodielectric measurements were carried out for the oxygen deficient sample. Resistivity measured in zero field and in ZFCH mode with 30 kOe field exhibiting semiconducting temperature dependence are shown  by the semilogarthmic plot in the inset of Fig. 4. The plots in both the measurements do not show any signature of $T_{\rm N}$ and $T_{\rm CO}$ in the temperature dependence. Magnetoresistance (MR) defined as [$\rho$ ($H$) - $\rho$ ($H$ = 0)]/$\rho$ ($H$ = 0) is found to be low $\sim$ 3 \% around $T_{\rm N}$. Besides, it does not reveal any signature around $T_{\rm N}$ or $T_{\rm CO}$ in the temperature dependence unlike the strongest MD effect in between 180 K and 200 K. Negative MR plotted with $H$ (MR-$H$ curve) is illustrated in Fig. 4 at 180 K, 210 K, and 150 K, being representative temperatures at $T_{\rm N}$, incommensurate charge ordered state, and commensurate charge ordered state, respectively. The arrow in Fig. 4 highlights the values of MR at 30 kOe which are nearly close ($\sim$ 3 \%), exhibiting a very small increasing trend of MR with decreasing temperature. We also measured the MR-$H$ curves at 180 K, 210 K, and 150 K for the oxygen annealed sample where the field dependence and the magnitude up to 30 kOe are almost similar like the oxygen deficient sample. The observed magnetoresistance results emphasize that the strongest MD effect in the range 180 - 200 K is not correlated with the MR. We further note another peak in the MD effect at $T_{\rm P}$ where a distinct peak in $\tan\delta$, ZFCH magnetization, and resistivity at 50 kOe in FCC mode is observed for the oxygen  deficient sample.

\begin{figure}[t]
\vskip 0.0 cm
\centering
\includegraphics[width = 8 cm]{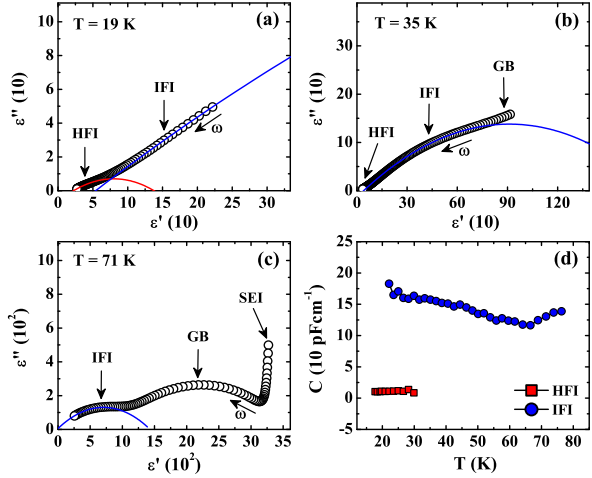}
\caption {Cole-Cole plots at 19 K (a), 35 K (b), and 71 K (c). (d) Estimated capacitances obtained from the high and intermediate frequency components as a function of temperature  for the oxygen deficient sample.} 
\label{Fig. 5}
\end{figure}
\begin{figure}[t]
\vskip 0.0 cm
\centering
\includegraphics[width = 8 cm]{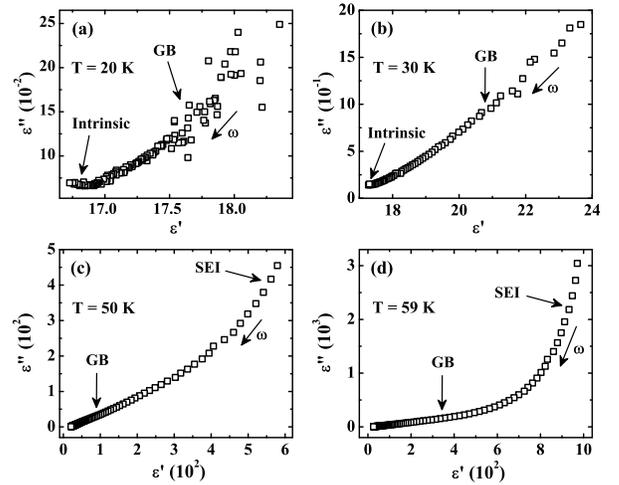}
\caption {Cole-Cole plots at 20 K (a), 30 K (b), 50 K (c), and 59 K (d) for the oxygen annealed sample.} 
\label{Fig. 6}
\end{figure}

Figure 5 shows the complex plane plots of $\epsilon$ (the Cole-Cole plots) at 19 K, 35 K, and 71 K. At 19 K an arc in the high frequency region defined as high-$f$ intrinsic (HFI) component is highlighted in Fig. 5(a). The high frequency component is shifted toward higher frequency with increasing temperature which could be analyzed satisfactorily until 30 K. In addition to the HFI component, another intrinsic component in the intermediate frequency range defined as intermediate-$f$ intrinsic (IFI) component is shown partly at 19 K and more clearly at 35 and 71 K in the Cole-Cole plots where the fits are indicated by the continuous lines in Figs. 5 (a), (b) and (c), respectively. At 71 K three arcs corresponding to IFI, grain boundary (GB), and sample-electrode interface (SEI) components are observed in order with decreasing frequency [Fig. 5(c)]. The values of the capacitances ascribed to the intrinsic components corresponding to high and intermediate frequency ranges as a function of temperature are shown in Fig. 5(d) where the IFI component lies in the range  110 - 180 pFcm$^{-1}$ and the HFI component around $\sim$ 10 pFcm$^{-1}$. The IFI component could be estimated unambiguously from the Cole-Cole plots until 77 K and the HFI cmponent until 30 K. Any distinct signature in the intrinsic dielectric response is not convincingly noticed around $T_{\rm P}$. We note that both the components (HFI and IFI) coexist at low temperature for the oxygen deficient sample whereas the HFI component is absent in the oxygen annealed sample. In Fig. 6 the Cole-Cole plots are depicted at 20 K, 30 K, 50 K, and 59 K for the oxygen annealed sample. At 59 K [Fig. 6(d)] the plot shows a sharply rising feature at the low-$f$ side (right hand side) which is the SEI component. It is followed by the GB component at the higher frequency side. At a nearby lower temperature, 50 K [Fig. 6(c)], the SEI component is on the verge of moving out of the frequency window. At a still lower temperature, 30 K [Fig. 6(b)], SEI component has completely moved out and the GB region is present. An intrinsic component seems to have just appeared, which is clearly visible in the plot of 20 K [Fig. 6(a)] in the high-$f$ region. The intrinsic component is prominently seen along with the consecutive GB region in Fig. 6(a). The high-$f$ intrinsic component is not observed for the oxygen annealed sample within the measured frequency window. The dielectric results are consistent with the magnetization and magnetoresistance results, suggesting that the disorder in the charge ordering introduced by the oxygen nonstoichiometry may lead to the phase coexistence below $T_{\rm P}$. The coexistence of high and low  frequency components in the dielectric measurement further  confirms the phase coexistence for the oxygen deficient sample. 

Magnetization, magnetoresistance, and dielectric results clearly demonstrate that the properties are modified substantially by the oxygen stoichiometry in the low temperature region for $T \leq T_{\rm P}$ whereas the properties remain almost unaltered in the  high temperature region. The origin of the appearance of $T_{\rm P}$ is needed to be elucidated further using microscopic experimental tool such as neutron diffraction studies in the oxygen  deficient sample. The observation of the strongest MD effect around the magnetic transition is rather appealing. Careful observation of magnetoresistance does not reveal any signature in the temperature dependence at the region around which the strongest magnetodielectric response is observed. In fact, several intricate processes such as paramagnetic to antiferromagnetic transition, IC to C charge and orbital ordering have been reported in the temperature region around which the strongest MD effect is noticed in the current investigation.  Extensive investigations on this  compound based on resonant x-ray scattering, \cite{zim} electron microscopy, \cite{mori,asa} NMR, \cite{yaku} neutron diffraction \cite{kaji} establish the microscopic views of intricate processes in the high temperature region. In addition to the charge ordering at $T_{\rm CO}$ an onset of orbital ordering ($T_{\rm OO}$) was also shown to be associated exactly at the same temperature where IC to C orbital ordering was reported around 215 K with an orbital ordering vector {\bf Q}=(0,1/2,0) in the commensurate orbital ordered state. \cite{kaji,asa} A ferromagnetic spin fluctuation involved with the orbital disordering in the IC state was pointed out from the neutron \cite{kaji} and NMR \cite{yaku} investigations in the temperature range between 215 K and $T_{\rm OO}$. The transition from IC to C charge ordering also takes place associated with paramagnetic to antiferromagnetic transition at 180 K which does not match with the IC to C orbital ordering temperature. Therefore, the compound undergoes a delicate interplay among charge, orbital, structure, and spin degrees of freedom in the temperature range between $T_{\rm N}$ and $T_{\rm CO}$ or $T_{\rm OO}$ around which a considerable magnetodielectric response is noticed. The strongest MD effect in between 180 K and 200 K is well within the range where IC to C charge and orbital ordering has been reported. The transformation from IC to C states associated with the charge and orbital ordering is suggested to be highly sensitive to the external perturbation such as magnetic field which may lead to the largest MD effect. The results are analogous to that of $R$Mn$_2$O$_5$ ($R$ = Tb, Dy, Ho) where colossal MD effect was suggested to be involved with the IC to C magnetic transition. \cite{cruz,blake,hur} We further note that a considerable MD effect at 280 K ($\sim$ 28 \%) is still observed well above $T_{\rm CO}$ at 245 K. Despite the fact that long range charge and orbital ordering occurs simultaneously at 245 K, the onset of short range orbital and charge ordering actually starts at much higher temperature than $T_{\rm CO}$ or $T_{\rm OO}$ which might be related with the considerable MD effect at 280 K. Large MD effect close to room temperature is significant which has a tremendous technological importance.

\section{Summary and conclusion}
The experimental observations based on magnetization, magnetoresistance, and magnetodielectric results demonstrate a marked difference in the low-$T$ state depending on the oxygen stoichiometry whereas the high-$T$ state remains almost unaltered. A disorder driven phase separation is observed at low temperature which is attributed to the oxygen deficiency. Interestingly, the signature of charge ordering temperature at $T_{\rm CO}$ = 245 K is clearly noticed in the field-dependence as well as field-cooled effect of  $\tan\delta$. A fairly large magnetodielectric effect observed in the measured temperature range is rather appealing. The considerable magnetodielectric effect ($\sim$ 28 \%) close to room temperature, measured with 30 kOe field and at 280 K, is fascinating for technological applications. The strongest magnetodielectric effect observed in between 180 K and 200 K is found to be uncorrelated with the magnetoresistance. In fact, the strongest magnetodielectric effect noticed in the current observation is in the temperature region where several intricate processes such as paramgnetic to antiferromagnetic transition, incommensurate to commensurate charge and orbital ordering have been reported. The results suggest a possible magnetoelectric coupling in the half doped charge ordered manganite, Pr$_{0.5}$Ca$_{0.5}$MnO$_3$, and the compound may be recognised as a new candidate in mutiferroics.

\noindent
{\bf Acknowledgement}
S.G. wishes to thank Department of Science and Technology, India (Project No. SR/S2/CMP-46/2003) for the financial support.  

\end{document}